\documentclass[epj]{svjour}
\usepackage{latexsym,amsfonts,amsmath,amssymb,bm}
\usepackage{graphics}

\begin{document}
\title{Why boundary conditions do not generally determine the universality class for boundary critical behavior}

\author{H.~W.\ Diehl}                   
%
%
\institute{Fakult\"at f\"ur Physik, Universit\"at Duisburg-Essen, D-47058 Duisburg, Germany}
\date{Received: date / Revised version: date}
%
\abstract{
Interacting field theories for systems with a free surface frequently exhibit distinct universality classes of boundary critical behaviors depending on gross surface properties.  The boundary condition satisfied by the continuum field theory on some scale may or may not be decisive for the universality class that applies. In many recent papers on boundary field theories, it is taken for granted that Dirichlet or Neumann boundary conditions decide whether the ordinary or special boundary universality class is observed.  While true in a certain sense for the Dirichlet boundary condition, this is not the case for the Neumann boundary condition. Building  on results that have been worked out in the 1980s, but have not always  been appropriately appreciated in the literature, the subtle role of boundary conditions and their scale dependence is elucidated and the question of whether or not they determine the observed boundary universality class is discussed.
} 
\titlerunning{Why  boundary conditions do not generally determine surface universality classes}
\maketitle
\section{Introduction}
\label{sec:intro}

In the middle of the 1970s, it became clear that systems with a free surface that exhibit bulk critical behavior  representative of a given bulk universality class may display distinct kinds of boundary critical behaviors \cite{LR75a,LR75b,BM77c,Bin83,Die86a,Die97}.\footnote{For reviews on boundary critical behavior and more complete lists of references, see \cite{Die82,Bin83,Die86a,Die97}.} In the 1980s and 1990s many powerful field-theory approaches based on dimensionality expansions about the upper and lower critical dimensions \cite{RG80,DD80,Ree81,DD81a,DD81b,DD83a,DN86}, the $1/N$~expansion \cite{BM77c,BM77a,OO83,OO83b,OO84},
 and the massive field-theory approach in fixed space dimensions \cite{DS94,DS98} were extended to systems with free surfaces. These made detailed quantitative analyses of appropriate field-theory models possible, which complemented studies using alternative methods such as position-space renormalization, high-temperature series expansions, and Monte Carlo simulations \cite{LB90a,RDWW93,PS98,Kre00,Ple04,DBN05,Has11,Has11b}.\footnote{The vast amount of published papers on the subject makes it impossible to include a complete list of citations. Our choice of cited papers on field-theory approaches and Monte Carlo results is motivated by what will be helpful below. The interested reader is advised to consult the review papers \cite{Bin83,Die86a,Die97} and \cite{Ple04} for further references. } A parallel development initiated by the seminal work \cite{BPZ84,BPZ84a} on the infinite conformal symmetry in two-dimensional quantum field theory was the extension and application of conformal field-theory methods to two-dimensional systems with boundaries (see \cite{Car84,DiFMS97,Gin90} for reviews).
 
 A more recent development has been the use of conformal bootstrap techniques to ($d>2$)-dimensional systems  \cite{PRV19} and their extension to systems with boundaries (see, e.g., \cite{LRR13,GLMR15,BHS19,KP20} and their references). These advances and the ongoing great interest in the AdS/CFT correspondence  have spurred  considerable recent activity by the high-energy theory community in the field of boundary conformal field theories \cite{HH17,HS19,Shp19,HK20,PS20,DHS20}. Using these techniques, the series expansions to second order in $\epsilon=4-d$ about the upper critical dimension  $d^*=4$  of the surface correlation exponents $\eta^{\text{ord}}_\|$ and $\eta^{\text{sp}}_\|$ at the ordinary and special transitions, which have been known since the beginning of the 1980s \cite{RG80,DD80,Ree81,DD81a,DD81b,DD83a}, were recovered \cite{BHS19,KP20}.\footnote{Unfortunately, the $O(\epsilon^2)$ result for the second independent surface exponent of the special transition, namely the crossover exponent $\Phi$ \cite{DD81b,DD83a,Die86a}, has so far not been determined by such bootstrap methods.} In much of the recent work addressing the high-energy physics community, as well as some older work in condensed matter physics, it is claimed or taken for granted that the choice of Dirichlet and Neumann boundary conditions for the field theory is the proper way of selecting and identifying the ordinary and special transitions, respectively. This overlooks the important fact that boundary conditions generally ought to be considered as scale-dependent properties. Clearly, in order that the boundary condition reflects a property of the surface universality class (ordinary or special), it must correspond to a fixed point, either infrared stable or unstable. This is the case for the Dirichlet boundary condition, which holds at the infrared-stable ordinary fixed point. By contrast, the Neumann boundary condition does not correspond to a fixed point, neither an unstable nor stable one, except in the free field case. Identifying the special transition via the Neumann boundary condition therefore is unacceptable, dangerous,  and misleading.
 
 The aim of the present paper is to discuss the identification of the surface universality classes and the role of boundary conditions in some detail. Let me stress that the reasoning presented below is based on arguments most of which can be found in the literature, notably in \cite{Die86a,Die97}. In fact, there exists at least one high-energy paper in which the difference between the Dirichlet and Neumann boundary conditions (to be elucidated below) is fully appreciated and its consequences for simulations of lattice models are pointed out  \cite{Lue06}. The continuing identification of the special transition with the Neumann boundary condition reveals a serious lack of understanding of the subtle issues  of  the scale dependence of boundary conditions and the identification of surface universality classes.  The explanations given below try to counterbalance this.
 
 The remainder of this paper is organized as follows. In Section~\ref{sec:models}, we briefly recall the familiar Ising lattice model with free surfaces, its mapping to a $\phi^4$ theory, and the resulting mesoscopic Robin boundary condition. We argue that the field theory one obtains from the Ising model in the case where its surface couplings are critically enhanced so that a special transition occurs must not be expected to satisfy Neumann boundary conditions. We also show that a mesoscopic Neumann boundary condition could well result  from an Ising model whose surface couplings are subcritically enhanced. In  Section~\ref{sec:fieldtheory}, we recollect some necessary background of the field theory approach to the semi-infinite $\phi^4$ theory and its renormalization group analysis. This is used to discuss the boundary conditions that hold asymptotically on large length scales, both for the ordinary and special fixed point. A  comparison with the situation in the case of the extraordinary transitions is also made. Section~\ref{sec:conc} briefly summarizes our findings and conclusions.

 \section{Lattice models and field theory}
\label{sec:models}
Consider a simple cubic lattice with sites $\bm{i}\in\mathbb{Z}^d$, consisting of $\mathcal{N}_z+1$ $(d-1)$-dimensional layers $[0,\mathcal{N}_y]^{d-1}$ labeled by $z=0,\ldots,\mathcal{N}_z$. To each site $\bm{i}$ a spin variable $s_{\bm{i}}=\pm1$ is attached. The spins interact via nearest-neighbor bonds $K_1=J_1/k_BT$ in the surface layers $z=0,\mathcal{N}_z$ and $K=J/k_BT$ elsewhere. Its configurational energy$/k_{B}T$ is given by\begin{equation}
\label{eq:Ismod}
\mathcal{H}_{\text{lat}}=K_1\sum_{\substack{\langle \bm{i},\bm{j}\rangle\\ z=0,\,\mathcal{N}_z}}s_{\bm{i}}s_{\bm{j}}+K\sum_{\substack{\langle \bm{i},\bm{j}\rangle\\0<z<\mathcal{N}_z}}s_{\bm{i}}s_{\bm{j}}.
\end{equation}

Along all directions $\bm{y}$ parallel to the surface layers periodic boundary conditions are chosen. Along the $z$-direction, free boundary conditions are used. We take the limits $\mathcal{N}_y\to\infty$ and $\mathcal{N}_z\to\infty$. 

This Ising model with free surfaces (and large but finite $\mathcal{N}_y$ and $\mathcal{N}_z$) has been frequently used in Monte Carlo simulations \cite{LB90a,RDWW93,PS98,Kre00,Ple04}. Its $(d{=}3)$-dimensional version is known to have a bulk critical  point at a critical coupling $K_{\text{c}}$  and to undergo an ordinary, special, and extraordinary surface transition depending on whether the ratio $r=K_1/K$ is smaller, equal, or larger than a critical value $r_{\text{sp}}$. These critical values \cite{LB90a,RDWW93,PS98,Ple04}
\begin{equation}
\label{eq:K1sp}
K_{\text{c}}\simeq 0.22,\qquad r_{\text{sp}}\equiv \left(K_1/K\right)^{\text{sp}}\simeq 1.5
\end{equation} depend, of course, on microscopic details. They  change if, say, a coupling $K_\perp\ne K$ is chosen between the surface layers and their neighboring layers. Likewise, they take different values for the Blume-Capel model used in the Monte Carlo analysis of \cite{Has11} to minimize corrections to scaling. 

To obtain a description of this model on mesoscopic length scales, one can coarse-grain to obtain a continuum field theory. An efficient and convenient, though admittedly approximate, way to reach this goal is to consider the mean-field theory for the lattice model~\eqref{eq:Ismod}, make a continuum approximation, and identify the field-theory action via the resulting Ginzburg-Landau (zero-loop) equations and by dropping contributions that may be expected to be irrelevant \cite{LR75b,Bin83,Die86a}. The resulting action is
\begin{eqnarray}
\label{eq:Hcont}
\mathcal{H}&=&\int d^{d-1}y\bigg\{\int_0^\infty dz\left[\frac{1}{2}(\nabla\phi)^2+\frac{\mathring{\tau}}{2}\phi^2+\frac{\mathring{u}}{4!}\phi^4\right]\nonumber\\
 &&\strut +\frac{\mathring{c}}{2}\phi^2\big|_{z=0}\bigg\},
\end{eqnarray}
where $\phi(\bm{y},z)$ is an order parameter field on the semi-infinite space $\mathbb{R}^{d-1}\times\mathbb{R}^+$, while $\mathring{\tau}$ and $\mathring{u}$ are bare bulk parameters. For the bare surface parameter $\mathring{c}$, the explained procedure yields the (approximate) value \cite{LR75b}
\begin{equation}
\label{eq:c0mf}
\mathring{c}=1-2(d-1)(K_1/K-1).
\end{equation}
Again, this value depends on microscopic details of the lattice model. Had we taken a coupling $K_\perp\ne K$ between the surface and its adjacent layer, a different value would result (see equation~(2.19) of \cite{Die86a}).

To make this field theory well-defined beyond zero-loop order, we need a regularization. For our purposes, it will be convenient and sufficient to use either dimensional regularization or a cutoff regularization in which the momentum integrations are restricted such that the momentum component $\bm{p}$ conjugate to $\bm{y}$ is restricted by $|\bm{p}|\le\Lambda$. The length scale $2\pi/\Lambda$ may be viewed as the linear block size to which one has coarse-grained. 

From the boundary terms of the classical equations of motion $\delta\mathcal{H}=0$, one obtains the Robin boundary condition \cite{LR75b,Bin83,Die86a}
\begin{equation}
\label{eq:bcbare}
\partial_n\phi=\mathring{c}\phi,
\end{equation}
where $\partial_n$ ($=\partial_z|_{z=0}$) denotes the derivative along the inner normal.
Just as the classical equations of motion, this boundary condition holds beyond  zero-loop order in an operator sense (inside of averages) \cite{Die86a,Die97}.
For $\mathring{c}=\infty$ and $\mathring{c}=0$, the boundary condition becomes a Dirichlet and Neumann boundary condition, respectively. Note that  $\mathring{c}=0$ 
is the zero-loop (``classical'') critical value $\mathring{c}_{\text{sp,cl}}$ of the special transition for $d>2$. From eq.~\eqref{eq:c0mf} we see that it corresponds to a surface enhancement ratio 
\begin{equation}
\label{eq:rN}
r_{\text{N}}=\frac{2d-1}{2(d-1)}\mathop{=}_{d=3}\frac{5}{4}.
\end{equation}
That the ${d=3}$ Monte Carlo value $r_{\text{sp}}$ for critical enhancement given in eq.~\eqref{eq:K1sp} is roughly 25\% higher than $r_{\text{N}}$ is not surprising since the Ginzburg-Landau (``classical'') theory, unlike Monte Carlo calculations, neglects fluctuations. 

What this tells us is that there is no reason to believe that the Ising model with a critical surface enhancement $r=r_{\text{sp}}$ maps onto the field theory~\eqref{eq:Hcont} with  Neumann boundary conditions. It should also be clear that if we do the mapping of the lattice model~\eqref{eq:Ismod} via a  more elaborate coarse-graining method by integrating out short length-scale degrees of freedom up to the minimal length scale $2\pi/\Lambda$ (following  Wilson \cite{WK74}), and obtain  a given value of $\mathring{c}$, then choosing another value $\Lambda'$ would give us a different value $\mathring{c}'$.

That the Neumann boundary condition is not a  property of the special transition can also be seen by computing $\mathring{c}_{\text{sp}}$ perturbatively for the cutoff regularized continuum model~\eqref{eq:Hcont}. Upon setting $\mathring{\tau}$ to its critical value $\mathring{\tau}_{\text{c}}$, one can compute $\mathring{c}_{\text{sp}}$ from the condition that the surface susceptibility $\chi_{11}=\int d^{d-1}y\, \langle\phi(\bm{y},0)\phi(\bm{0},0)\rangle$ diverges, i.e., $[\chi_{11}(\mathring{\tau}_{\text{c}},\mathring{c}_{\text{sp}})]^{-1}=0$. The result for $d=4-\epsilon>2$,
\begin{equation}
\label{eq:cspbare}
\mathring{c}_{\text{sp}}=-\frac{\mathring{u}}{8}K_{d-1}\Lambda\mathring{u}/\Lambda^\epsilon+O(\mathring{u}^2)
\end{equation}
with
\begin{equation}\label{eq:Kd}
K_d\equiv 2^{1-d}\pi^{-d/2}/\Gamma(d/2)
\end{equation}
can be gleaned from equation~(3.95) of \cite{Die86a}.

What we have learned above about the cutoff dependence of $\mathring{c}$ and hence the boundary condition for the bare theory obviously applies also for mappings of the Ising model~\eqref{eq:Ismod} with subcritical ($r<r_{\text{sp}}$) or supercritical enhancements ($r>r_{\text{sp}}$). Nevertheless, there are important differences between the cases of Dirichlet and Neumann boundary conditions. Before we can discuss these, we  must first  recollect in the next section some well-known facts about the renormalization group analysis of the model~\eqref{eq:Hcont}. 

 \section{Renormalization}
 Let   \begin{equation}
 \label{eq:GNM}
 G^{(N,M)}=\bigg\langle\bigg[\prod_{i=1}^N\phi(\bm{x}_i)\bigg]\bigg[\prod_{j=1}^M\phi|_s(\bm{y}_j)\bigg]\bigg\rangle^{\text{cum}}
 \end{equation}
 denote  the cumulants (connected ($N+M$)-point Green's functions) associated with the action~\eqref{eq:Hcont}. They
 involve $N$ fields $\phi$ at positions $\bm{x}_i$ away from the boundary and $M$ fields $\phi|_s(\bm{y}_j)=\phi(\bm{y_j},0)$ at points $\bm{y}_j$ on the surface $z=0$. Their ultraviolet (UV) singularities can be absorbed through the reparametrizations
\label{sec:fieldtheory}
\begin{equation}\label{eq:bulkrep}
\phi=Z_\phi^{1/2}\phi^{\text{ren}}, \quad
\mathring{\tau}=\mu^2Z_\tau\tau+\mathring{\tau}_{\text{c}},\quad
\mathring{u}N_d=\mu^{4-d} Z_uu,
\end{equation}
and
\begin{equation}\label{eq:surfrep}
\mathring{c}=\mu Z_c\, c+\mathring{c}_{\text{sp}}\qquad
\phi |_s=(Z_\phi Z_1)^{1/2}\phi|^{\text{ren}}_s.
\end{equation}
Here, $\mu$ is a momentum scale, and $N_d$ is a convenient normalization constant that is absorbed in $u$. The latter was chosen as $(4\pi)^{-d/2}$ in \cite{DD80,DD81a,DD81b,DD83a,Die86a}; the alternative choice $ N_d=2\,(4\pi)^{-d/2}\Gamma(3-d/2)/(d-2)$ made elsewhere \cite{GD08,DS11} is still compatible with the explicit two-loop expressions given in the former references. The counterterms $\mathring{\tau}_{\text{c}}$ and $\mathring{c}_{\text{sp}}$ absorb UV singularities quadratic and linear in the cutoff, respectively, in the cutoff regularized theory. The critical value of the renormalized surface enhancement variable $c$ for which the special transition occurs is $c_{\text{sp}}=0$.

Let us also recall that in a perturbative approach based on the  dimensionally regularized theory and the $\epsilon$ expansion, the shift $\mathring{c}_{\text{sp}}$ vanishes, as does $\mathring{\tau}_{\text{c}}$.\footnote{One can benefit from this in a  perturbative approach based on dimensional regularization and the $\epsilon$ expansion by performing calculations directly at $\tau=c=0$, provided one carefully avoids undefined infrared singular quantities.} However, beyond perturbation theory these quantities are known to be nonzero and of the form $\mathring{\tau}_{\text{c}}=\mathring{u}^{2/\epsilon}\mathcal{T}(\epsilon)$ and $\mathring{c}_{\text{sp}}=\mathring{u}^{1/\epsilon}\mathcal{C}(\epsilon)$ in the dimensionally regularized theory, where the functions $\mathcal{T}(\epsilon)$ and $\mathcal{C}(\epsilon)$ have poles at $\epsilon_k=2/k,\,k\in \mathbb{N}$ \cite{Sym73b,DS94,DS98}.

Under the renormalization group flow implied by a change $\mu\to\mu\ell$ of the momentum scale the enhancement variable $c$ is mapped onto  the scale-dependent  value $\bar{c}(\ell)$ which varies as
\begin{equation}
\label{eq:cbar}
\bar{c}(\ell)\sim \ell^{-y_c}c
\end{equation}
in the infrared limit $\ell\to 0$. The RG eigenvalue $y_c$ can be expressed as $y_c=\Phi/\nu$ in terms of the surface crossover exponent $\Phi$ associated with the special transition and the bulk correlation-length exponent $\nu$ \cite{DD81b}. The $\epsilon$ expansions  of $y_c$ and $\Phi$ to $O(\epsilon^2)$ can be found in \cite{Die86a}, along with estimates  of the corresponding values for $d=3$. Improved ${d=3}$  field-theory estimates of these exponents, which additionally take into account the two-loop results of the massive field theory approach \cite{DS94,DS98}, are given in \cite{DS98}. For $\Phi$ at $d=3$, the estimate found there is $\Phi({d=3})\simeq 0.54$; the most recent  Monte Carlo calculations \cite{RDWW93,PS98,DBN05,Has11} give somewhat lower values $\simeq 0.45$. Unfortunately, the estimates  obtained via the conformal bootstrap approach  for the exponents of the special transition at $d=3$ are not yet competitive, unlike those for the ordinary transition \cite{GLMR15}.

For  initial values $c>0$ and $c<0$, the running variable $\bar{c}(\ell)$ is driven to the fixed-point values $c^*_{\text{ord}}=\infty$ and $c^*_{\text{ex}}=-\infty$ of the ordinary and extraordinary fixed points. At the ordinary fixed point, the field $\phi$ and hence the cumulants $G^{(N,M)}$ satisfy Dirichlet boundary conditions (which means in particular that they vanish when $M>0$). A nice way to confirm this is the use of the boundary operator expansion (BOE) \cite{DD81a,Die86a,Die97}. For the case of the  ordinary fixed point, the leading contribution of the BOE of $\phi$ originates from $\partial_n\phi$. To this end, one can set $\mathring{c}=\infty$ and consider the $\mathring{c}=\infty$ analogs of the cumulants~\eqref{eq:GNM} involving $M$ surface operators $\partial_n\phi$ rather than $\phi|_s$. The renormalization of $\partial_n\phi=(Z_\phi Z_{1,\infty})^{1/2}[\partial_n\phi]^{\text{ren}}$ involves another  renormalization factor $Z_{1,\infty}$, which was determined to two-loop order in \cite{DD80,DD81a}. Let me just quote the near-boundary behavior one obtains for both the ordinary and the special fixed points by including  the leading contribution to the BOE; one finds \cite{DD81a,DD83a,Die86a}
\begin{equation}
\label{eq:sdb}
\phi^{\text{ren}}(\bm{y},z)\mathop{\sim}_{z\to 0} \begin{cases}z^{\varsigma^{\text{ord}}}\,[\partial_n\phi]^{\text{ren}}(\bm{y})&\text{for }c=\infty,\\
z^{-\varsigma^{\text{sp}}}\,\phi|_s^{\text{ren}}(\bm{y})&\text{for }c=0,
\end{cases}
\end{equation}
where the exponents $\varsigma^{\text{ord/sp}}$ can be expressed as
\begin{equation}
\label{eq:sigmas}
\varsigma^{\text{ord}}=(\beta_1^{\text{ord}}-\beta)/\nu=\eta_\perp^{\text{ord}}-\eta
\end{equation}
and
\begin{equation}
\varsigma^{\text{sp}}=(\beta-\beta_1^{\text{sp}})/\nu=\eta-\eta_\perp^{\text{sp}}
\end{equation}
in terms of the bulk magnetization exponent $\beta$ and the surface magnetization exponents $\beta_1^{\text{ord/sp}}$ or the surface correlation exponents  $\eta_\perp^{\text{ord/sp}}$ of the ordinary and special transitions.

Note that field-theory estimates \cite{DD80,Ree81,DD81a,DD81b,DD83a,Die86a,DS98}, Monte Carlo results \cite{LB90a,RDWW93,PS98,DBN05,Has11,Has11b}, and conformal bootstrap approaches consistently show that both exponents $\varsigma^{\text{ord/sp}}$ are positive for the $d=3$ Ising case.\footnote{From the field-theory estimates of \cite{DS98}, the Monte Carlo results \cite{Has11,Has11b}, and the conformal bootstrap approach \cite{GLMR15} one obtains approximately the same value $\varsigma^{\text{ord}}\simeq 0.75$. Likewise, one finds from  \cite{Has11,Has11b} and \cite{DS98} the values $\varsigma^{\text{sp}}\simeq 0.16$ and $0.1$, respectively. We refrain from giving an estimate for $\varsigma^{\text{sp}}$ resulting from the bootstrap approach of \cite{GLMR15}  because the results for the exponents of the special transition are not very precise according to the authors' own judgment. } In fact, it can be rigorously shown by means of Griffiths inequalities \cite{Gri67a,KS68} that $\varsigma^{\text{ord}}\ge 0$. To see this, consider two semi-infinite ${(d=3})$-dimensional Ising models whose spins are coupled via nearest-neighbor bonds of equal strengths $K\equiv K_1$. Assume that they consist of two-dimensional layers at $z=0,1,\ldots$ and $z=-1,-2,\ldots$, respectively. Upon  coupling their surface layers via nearest-neighbor bonds of strength $K_\perp$, we can compare the magnetizations \sloppy $m_{0,K_\perp}\equiv\sum_{i \text{ with }z=0}\langle s_i\rangle/\sum_{i \text{ with }z=0}1$ per site of the layer $z=0$ for the cases $K_\perp=0$ and $K_\perp=K$. Since for $K_\perp=K$ we have additional ferromagnetic interactions, the Griffiths inequalities tell us that  $m_{0,K}\ge m_{0,0}$. However, these magnetizations vary as $m_{0,0}\sim|t|^\beta$ and $m_{0,K}\sim |t|^{\beta_1^{\text{ord}}}$ as $t\equiv (T-T_{\text{c}})/T_{\text{c}}\uparrow 0$, where $T_{\text{c}}$ is the bulk critical temperature. It follows that $\beta_1^{\text{ord}}\ge \beta$ because otherwise the inequality $m_{0,K}\ge m_{0,0}$ would be violated for sufficiently small $|t|$.

The mean-field (zero-loop) result for $\varsigma^{\text{ord}}$ is $1$. Its $\epsilon$ expansion
\begin{equation}
\label{eq:sigmaordeps}
\varsigma^{\text{ord}}=1-\frac{ 57 \epsilon +324}{1944}\epsilon +O(\epsilon^3)
\end{equation}
follows in a straightforward fashion from the known $O(\epsilon^2)$ expressions for the exponents in eq.~\eqref{eq:sigmas},
and from the exact $({d=2)}$-Ising-model results $\eta_\perp=5/8$ and $\eta=1/4$ (see, e.g., \cite{Car84}) we get $\varsigma^{\text{ord}}({d=2})=3/8$. Hence $\varsigma^{\text{ord}}$ may be expected to be strictly positive for all $d\ge 2$. Thus on scales large compared to microscopic ones (lattice spacing $a$), the order parameter field $\phi$ satisfies indeed a Dirichlet boundary condition at the surface. 

This should be contrasted with the situation at the special fixed point. The exponent $\varsigma^{\text{sp}}$ vanishes in mean-field theory (and hence for $d\ge 4$)  and has the $\epsilon$~expansion
\begin{equation}
\label{eq:sigmaspeps}
\varsigma^{\text{sp}}=\frac{1}{162} (27-\epsilon ) \epsilon+O(\epsilon^3).
\end{equation}
The $O(\epsilon^2)$ expression  increases monotonically from zero at $d=4$ to $13/81=0.168\ldots$ at $d=3$. Thus,  $\varsigma^{\text{sp}}>0$ for $3\le d<4$ so that the order parameter, extrapolated from scales $\gg a$ to small distances $z$,  \emph{diverges} in the limit $z\to 0$. Only at $d\ge 4$ or in the free field case does it satisfy a Neumann boundary condition.

This is not the place for a detailed discussion of the extraordinary transition. However, it may be helpful to recall for comparison the short-distance behavior at the extraordinary fixed point $c=-\infty$. Since  $\phi(\bm{y},z)$ has the anomalous dimension $\beta/\nu$ and  $\langle\phi(\bm{y},z)\rangle$  does not vanish at $T_{\text{c}}$ in the case of the extraordinary transition (i.e., for supercritical enhancement $c<0)$), one concludes that $\langle\phi(\bm{y},z\rangle$ decays  at $T_{\text{c}}$  as $z^{-\beta/\nu}$ for large $z$ \cite{BM77c,OO84,Die86a,Shp19,DHS20,DS93}.\footnote{This term evidently results from the contribution of the $1$ operator to the BOE. The contribution from the $zz$ component $T_{zz}(\bm{y})$ of the stress-energy tensor  \cite{Die97} yields the decays  $\langle\phi(\bm{y},z)\phi(\bm{0},z)\rangle^{\text{cum}}\mathop{\sim}\limits_{y\to\infty}
 y^{-(d-2+\eta_\|^{\text{ex}})}\sim y^{-2d}$ and $\langle\phi(\bm{y},z)\phi(\bm{y},z')\rangle^{\text{cum}}\mathop{\sim}\limits_{z\to\infty}z^{-(d-2+\eta_\perp^{\text{ex}})}\sim z^{-(3d-2+\eta)/2}$, giving $\eta_\|^{\text{ex}}=d+2$ and $\eta_\perp^{\text{ex}}=(d+2+\eta)/2$. } Extrapolated to $z\to 0$, this means that it diverges upon approaching the surface. This holds even in Ginzburg-Landau theory and hence for $d>4$.

 \section{Conclusions}
\label{sec:conc}

The purpose of this paper has been to explain why except in the free-field case the Neumann boundary condition \emph{is not a property of the special transition}, as frequently taken for granted in the literature, where it tends to get used for the identification of the special transition. It is essential to realize that the boundary condition one has to deal with in a field-theory description is a scale-dependent property.  Since in  microscopic models the spin variables fluctuate, coarse graining to a mesoscopic minimal length scale $2\pi/\Lambda$  yields a value for the bare surface enhancement variable $\mathring{c}$  appearing in the mesoscopic Robin boundary condition~\eqref{eq:bcbare} that depends on the choice of $\Lambda$. 

In section~\ref{sec:models}, we discussed the mapping of a simple Ising model with a free surface and modified surface bonds onto  a  semi-infinite $\phi^4$ model. We saw that the ratio $r_{\text{N}}=(K_1/K)_{\text{N}}$ of the surface and bulk couplings $K_1$ and $K$ that yield $\mathring{c}=0$ according to the approximate relation~\eqref{eq:c0mf} --- and hence a mesoscopic Neumann boundary condition --- was considerably lower than the critical value $r=r_{\text{sp}}$ corresponding to the special transition. This means on the one hand that a Neumann boundary condition must not be expected to hold for the  field theory describing the special transition. On the other hand, it tells us that a mesoscopic Neumann boundary condition of the field theory, i.e., a value $\mathring{c}=0$, is likely to lie in the basin of attraction of the fixed point of the ordinary transition. In other words, even if one has a Neumann boundary condition  on a mesoscopic scale, the large-scale behavior may well be governed by the ordinary fixed point $c^*_{\text{ord}}=\infty$ and hence exhibit asymptotic Dirichlet boundary conditions. If so, the surface critical behavior of the ordinary transition is observed. Likewise, universal quantities that differ for the special and ordinary surface universality classes would display the value of the ordinary one. One interesting example, which has been much studied in recent years, is the fluctuation-induced force (``critical Casimir force'')  in a slab $\mathbb{R}^{d-1}\times[0,L]$ of thickness $L$  (see, e.g., \cite{KD91,KD92a,Kre94,SD08,DS11} and their references). At bulk criticality, this force decays as $\Delta_{\text{C}}L^{-(d-1)}$ for large thickness $L$, where the  Casimir amplitude $\Delta_{\text{C}}$ is a universal quantity that  depends on the bulk universality class and the two surface universality classes pertaining to the two semi-infinite systems bounded by the surface $z=0$ and $z=L$, respectively. If the values $c_1$ and $c_2$ of $c$ for both surfaces are subcritical, the amplitude takes the value $\Delta^{\text{ord,ord}}_{\text{C}}$ corresponding to large-scale Dirichlet boundary conditions and the fixed point $c^*_1=c^*_2=\infty$.

For the Dirichlet boundary condition to hold on large length scales it is sufficient that the surface enhancement variable $\mathring{c}$ appearing in the mesoscopic Robin boundary condition~\eqref{eq:bcbare} is subcritical ($\mathring{c}>\mathring{c}_{\text{sp}}$). This ensures that the initial value of the renormalized variable $c$ belongs to the basin of attraction of the ordinary fixed point.\footnote{This is why L\"uscher \cite{Lue06} calls Dirichlet boundary conditions ``natural.''} By contrast,  Neumann boundary conditions  satisfied by the field theory with a given minimal length $2\pi/\Lambda$ do not correspond to a fixed point of the renormalization group, neither a stable nor an unstable one (except in the free field case). Coarse graining to a larger minimal length scale would lead to modified (Robin) boundary conditions. 

Furthermore, it is not difficult to see that  boundary conditions are scale-dependent properties even on large scales. Consider a quantity $Q(z)$ such as \sloppy $\int_{\mathbb{R}^{d-1}} d^{d-1}y'\int_0^\infty dz'\langle\phi(\bm{y},z) \phi(\bm{y}',z')\rangle^{\text{cum}}$ that does not vanish at $T_{\text{c}}$ and assume that $c$ is subcritical, though nearly critical, i.e., positive but small. Associated with $c$ then is a length $\xi_c\sim c^{-1/y_c}$ that is much larger than the microscopic  one (lattice constant $a$). Clearly, depending on whether the distance $z$ satisfies $a\lesssim z\lesssim \xi_c$ or $z\gtrsim \xi_c \gg a$ the quantity $Q(z)$ must display the short-distance behaviors $\sim z^{-\varsigma^{\text{sp}}}$ and $\sim z^{\varsigma^{\text{ord}}}$ specified in eq.~\eqref{eq:sdb}. As $z$ decreases, the latter asymptotic behavior characteristic of the ordinary transition must smoothly cross over to that of the special transition. (For an illustration of this crossover behavior, see figure~22 of \cite{Die86a}.) Upon  equating the derivative $Q'(z)$  to zero, we can identify a distance $z_{\text{N}}(c)$ at which $Q(z)$ has a horizontal slope, i.e., satisfies  a Neumann boundary condition.
\section{Acknowledgments}
I would like to thank Tobias Hansen, Christoper Herzog, Miguel Paulos, and Mykola Shpot for informative correspondence and Sergei Rutkevich and Mykola Shpot for  a critical reading of the manuscript.%

%

\end{document}